

\def\Z{{\hbox{{\rm Z}\kern-.4em\hbox{\rm Z}}}}
\def\R{{\hbox{{\rm I}\kern-.4em\hbox{\rm R}}}}

\tolerance=10000
\input phyzzx

 \def\unit{\hbox to 3.3pt{\hskip1.3pt \vrule height 7pt width .4pt \hskip.7pt
\vrule height 7.85pt width .4pt \kern-2.4pt
\hrulefill \kern-3pt
\raise 4pt\hbox{\char'40}}}
\def\II{{\unit}}
\def\cM {{\cal{M}}}
\def\N {{\cal{N}}}
\def\F {{\cal{F}}}
\def\gij{g_{ij}}
\def\fii{\phi^i}
\def\fj{\phi^j}
\def\fk{\phi^k}

\def\nup#1({Nucl.\ Phys.\  {\bf B#1}\ (}
\def\Im{{\rm Im ~}}

\REF\sigmodsymanom{G. Moore and P. Nelson, Phys. Rev. Lett. {\bf  53} (1984)
117.}
\REF\MonOl{C. Montonen and D. Olive, Phys. Lett. {\bf B72} (1977) 117; E.
Witten and
D. Olive, Phys.
Lett. {\bf B78} (1978) 97.}
\REF\Sen {A. Sen, Nucl. Phys. {\bf B404} (1993) 109; Phys. Lett. {\bf 303B}
(1993); Int. J. Mod. Phys. {\bf A8} (1993) 5079; Mod. Phys. Lett. {\bf
A8} (1993) 2023; Phys.Lett. {\bf B329}, (1994) 217.}
\REF\sdual{A. Font, L. Ibanez, D. Lust and F. Quevedo, Phys. Lett. {\bf B249}
(1990) 35; S.J. Rey, Phys. Rev. {\bf D43}  (1991) 526.}
\REF\SS {J.H. Schwarz and A. Sen, Nucl. Phys. {\bf B411} (1994) 35; Phys.
Lett. {\bf 312B} (1993) 105.}
\REF\TD {A. Giveon, M. Porrati and E. Rabinovici, Phys. Rep. {\bf 244}
(1994) 77.}
\REF\DL {M.J. Duff and J.X. Lu, Nucl. Phys. {\bf B354} (1991) 141;
Phys. Lett. {\bf 273B} (1991) 409.}
\REF\DLb {M.J. Duff and J.X. Lu, Nucl. Phys. {\bf B416} (1993)  301.}
\REF\HT{C.M. Hull and P.K. Townsend, {\it Unity of string dualities},
        QMW preprint QMW-94-30, DAMTP preprint
        R/94/33, hepth/9410167.}
\REF\gencouplN{B. de Wit, P.G. Lauwers, R. Philippe, Su S.-Q. and
A. Van Proeyen, Phys. Lett. {\bf 134B} (1984) 37;
B. de Wit, P.G. Lauwers and A. Van Proeyen, Nucl. Phys. {\bf B255} (1985)
569.}
\REF\dWVP{B.\ de Wit and A.\ Van Proeyen, Nucl. Phys. {\bf B245}  (1984) 89.}
\REF\CecFerGir{S. Cecotti, S. Ferrara and L. Girardello,
Int. J. Mod. Phys. {\bf A4} (1989) 2457.}
\REF\WVP{B.\ de Wit, F.\ Vanderseypen and A.\ Van Proeyen, Nucl. Phys. {\bf
B400}
(1993) 463.}
\REF\CDAFVP{A. Ceresole, R. D'Auria, S. Ferrara and A. Van Proeyen,
{\it Duality Transformations in Supersymmetric Yang-Mills Theories coupled to
Supergravity},
preprint  CERN-TH 7547/94, POLFIS-TH. 01/95, UCLA 94/TEP/45,
KUL-TF-95/4, hep-th/9502072.}
\REF\dual{ S. Ferrara, J. Scherk and B. Zumino, Nucl. Phys. {\bf B121}
(1977) 393;
E. Cremmer, J. Scherk and S. Ferrara, Phys. Lett. {\bf 74B} (1978) 61;
B. de Wit,  Nucl. Phys. {\bf B158} (1979) 189;
E. Cremmer and B. Julia, Nucl. Phys. {\bf B159} (1979) 141.}
\REF\GZ {M.K. Gaillard and B. Zumino, Nucl. Phys. {\bf B193} (1981) 221.}
\REF\HW{C.M. Hull and E. Witten, Phys.Lett. {\bf 160B} (1985)  398.}
\REF\GNY{M.B. Green and D. Nemeschansky, in preparation.}
\REF\GH{M.B. Green and C.M. Hull, in preparation.}


\Pubnum{ \vbox{ \hbox {QMW-95-8} \hbox{KUL-TF-95/7}
\hbox{hep-th/9503022}} }
\pubtype{}
\date{March, 1995}

\titlepage

\title {\bf  PSEUDO-DUALITY}

\author{C.M. Hull}
\address{Physics Department,
Queen Mary and Westfield College,
\break
Mile End Road, London E1 4NS, U.K.}
\andauthor{A. Van Proeyen\foot{Onderzoeksleider,
NFWO, Belgium}}
\address{Instituut voor Theoretische Fysica, K.U. Leuven
\break
 Celestijnenlaan 200 D, B--3001 Leuven, Belgium}
\vskip 0.5cm

\abstract {Proper symmetries act on fields while pseudo-symmetries act on both
fields and coupling
constants. We identify the pseudo-duality groups that act as symmetries of the
equations of motion
of general systems of scalar and vector fields and apply our results to $N=2,4$
and $8$
supergravity theories. We present evidence that the pseudo-duality group for
both the heterotic and
type II strings toroidally compactified to four dimensions is $Sp(56;\Z)\times
D$, where $D$ is a
certain subgroup of the diffeomorphism group of the scalar field target space.
This contains the
conjectured heterotic $S\times T$ or type II $U$ proper duality group  as a
subgroup.}

\endpage

\chapter{Introduction}

Symmetries of field theories can be divided into two types. Proper symmetries
act on the fields
 and are associated with conserved Noether charges. There are also symmetries
which act  both on
fields and on coupling constants; these have no Noether charge and we shall
refer to them as {\it
pseudo-symmetries}. Sigma-model symmetries [\sigmodsymanom], reviewed below,
are
pseudo-symmetries. Another
example is the Montonen-Olive duality [\MonOl,\Sen] of $N=4$ supersymmetric
Yang-Mills
theory, in which
$SL(2,\Z)$ acts both on the fields and on the coupling constants $g, \theta$.
When $N=4$
supersymmetric Yang-Mills theory is embedded in the four-dimensional
toroidally
compactified
heterotic superstring, the coupling constants $g, \theta$ arise as expectation
values of the
dilaton and axion fields   and the conjectured $SL(2,\Z)$ S-duality of the
heterotic
superstring [\Sen,\sdual,\SS]
is a proper
symmetry acting on fields  (including the
dilaton and axion) only. Duality symmetries of supergravity and superstring
theories are proper
symmetries in this sense. For example, the heterotic superstring is believed
to
have
four-dimensional toroidally compactified $SL(2,\Z)\times O(6,22;\Z)$  $S\times
T$-duality [\SS-\TD]
while the
four-dimensional toroidally compactified type II superstring is believed to
have $E_7(\Z)$
U-duality [\HT]. However, $N=2$ supergravity coupled to $N=2 $ supermatter
[\gencouplN] has in general an
$Sp(2n;\R)$ duality symmetry which is in general a pseudo-symmetry,
of which only a subgroup (the scalar field target space isometry group) is a
proper symmetry group
[\dWVP -\CDAFVP]; this $Sp(2n; \R)$
pseudo-symmetry is broken to $Sp(2n; \Z)$ by quantum corrections.

This raises the possibility that supergravity and superstring theories can
have
pseudo-duality
symmetries above and beyond their known proper duality symmetries.  The
purpose
of this paper is to
investigate  a class of theories which includes supergravity and
superstring theories and show that in many cases there are indeed extra
pseudo-duality symmetries.
We will consider the general situation of actions containing
scalars coupled to $n$ Abelian vector fields,
which have been known for some time, at least in certain cases with $N=2$
supersymmetry
[\CecFerGir -\CDAFVP], to have an
$Sp(2n;\R)$ pseudo-duality symmetry, which contains the proper duality group
[\dual,\GZ] as a
subgroup; it was the investigation of this $Sp(2n;\R)$  symmetry that led to
the present work.
 We will
 show that in
general the  pseudo-duality symmetry group is in fact the much larger group
$Sp(2n;\R)\times
Diff({\cal M})$, where $Diff({\cal M})$ are the diffeomorphisms of
the target manifold of the scalars.
The previously studied $Sp(2n;\R)$ pseudo-duality symmetry [\CecFerGir
-\CDAFVP] occurs as a diagonal
subgroup of
$Sp(2n;\R)\times
Diff({\cal M})$.
Duality symmetries of superstring theories should, for compactifications that
maintain sufficient
space-time supersymmetry, be seen as duality symmetries of the low-energy
effective supergravity
theory, so that the existence of supergravity duality symmetry is a necessary,
but not sufficient,
condition for the existence of superstring duality [\Sen,\SS,\HT].

In supergravity
theories, duality is determined by its action on
the scalar and vector fields. It is usually  possible  to choose variables (by
a field redefinition) so that the fermions   are duality invariant
[\CDAFVP]. It is then necessary to check that the fermionic   terms in the
action are
invariant, which is indeed the case in all supergravity theories.
We shall consider here duality
symmetries of a general
scalar-vector system and then discuss the implications for supergravity and
superstrings.

\chapter{Sigma-Model Symmetries}

Consider the non-linear sigma-model in $D$ space-time dimensions
$$S= {1 \over 2} \int d^Dx \, \sqrt {-h} \, \sqrt {g}\, \gij
\partial _ \mu \phi ^i\partial ^\mu \phi ^j\ ,
\eqn\sig$$
where $x^\mu$ ($\mu = 0,1,  \dots , D-1$) are the coordinates of a space-time
with metric $h_{\mu
\nu}$ and the $\fii(x)$ ($i,j=1,\dots , d$) are scalar fields taking values in
a $d$-dimensional
target space {\cal M}; the $\fii$ can be thought of as coordinates for  {\cal
M}.
The quantity $\gij (\phi)$ is a metric on {\cal M}, but from the field-theory
point of view
represents a (generally infinite) set of coupling constants. For example, if
$\gij (\phi)$ has a
Taylor series expansion about $\fii =0$
$$
\gij (\phi)=g_{ij} ^{(0)} +
g_{ij,k} ^{(1)}\phi^k   +
g_{ij,kl} ^{(2)}\phi^k \phi^l  +
g_{ij,klm} ^{(3)}\phi^k\phi^l\phi^m+\dots
\eqn\tay$$
for some constants
$g_{ij,k_1 \dots k_n} ^{(n)}$,
then substituting \tay\ into \sig, the lagrangian takes the form $\sum g^{(n)}
\phi^n \partial \phi
\partial \phi$, so that
$g_{ij,k_1 \dots k_n} ^{(n)}$ is the coupling constant for a $\phi^n \partial
\phi
\partial \phi$ interaction.

Under a diffeomorphism of {\cal M},
$$ \fii \to \tilde \fii (\phi), \qquad \gij \to \tilde \gij
\eqn\diff$$
where
$$\tilde \gij (\tilde \phi(\phi)) {\partial\tilde \phi ^ i\over \partial  \phi
^k}
{\partial\tilde \phi ^j \over \partial  \phi ^l}= g_{kl} (\phi)
\eqn\gtis$$
The action \sig\ is invariant under \diff, which  consists of a field
transformation $ \fii \to \tilde \fii$ and a transformation
$\gij (\phi) \to \tilde \gij (\tilde\phi)$, which corresponds to a change
of the coupling constants, $g^{(n)} \to \tilde  g^{(n)}$. This is then
  a pseudo-symmetry, sometimes referred to as a sigma-model symmetry
[\sigmodsymanom].
Thus the pseudo-symmetry group of the non-linear sigma-model is the
diffeomorphism group $Diff({\cal
M})$.

It may be the case that there are special diffeomorphisms of {\cal M} which
leave the metric
invariant, i.e. for which
$$\tilde \gij(\phi) =\gij(\phi) \ .
\eqn\iso$$
Such diffeomorphisms are  isometries and are generated by Killing vector
fields.
They leave the metric and hence the coupling constants $ g^{(n)}$ invariant,
so
that they
constitute proper symmetries of the theory. The isometry symmetry generated by
a Killing vector
field $k^i(\phi)$ has a conserved current
$$J_\mu =k_i \partial _\mu \fii
\eqn\curr$$
and a corresponding Noether charge. Thus the group of proper
symmetries of the sigma-model is the isometry group $Iso({\cal M})$.

String theory in a background space-time {\cal M} can be described by a
two-dimensional non-linear
sigma-model of the form \sig\ with target space  {\cal M}, where $x^\mu$,
$\mu=0,1$, are viewed as
world-sheet coordinates and $\fii$ as space-time coordinates. The space-time
diffeomorphisms $Diff({\cal M})$
are a
pseudo-symmetry of this world-sheet theory, which leads to a first-quantized
formalism. However, in
a second-quantized formalism such as string field theory, from which general
relativity emerges in a
certain limit, the background metric $\gij$ is the vacuum expectation value of
the space-time
gravitational field, so that the    diffeomorphisms become a proper symmetry
of
the space-time field
theory. By contrast,
in a space-time sigma-model in which $x^\mu$ are space-time coordinates and
the
$\fii$ are
(second-quantized) space-time fields, the target space diffeomorphisms are a
pseudo-symmetry.
However,  $Diff({\cal
M})$ might become a proper symmetry of some ``third-quantized" theory in which
$\gij$ emerges as some
type of expectation value.

Sigma-model diffeomorphism symmetries are then pseudo-symmetries in general,
although the
isometry subgroup constitute proper symmetries. However, for world-sheet
sigma-models the
diffeomorphism pseudo-symmetry corresponds to a proper symmetry of the
space-time theory.

\chapter{Vector Field Duality}

Consider the four-dimensional Lagrangian
$$
L= -{1\over4}\sqrt {-h}\, m_{IJ}(\phi) F^{\mu\nu}F_{\mu\nu}^J +
{1\over8} \varepsilon^{\mu\nu\rho\sigma}a_{IJ}(\phi)F_{\mu\nu}^I
F_{\rho\sigma}^J
\eqn\aone
$$
for
$k$ abelian
vector fields  $A_\mu^I$ with field strengths $F_{\mu\nu}^I$ with couplings to
scalars $\phi^i$ through
the scalar
functions $m_{IJ}(\phi)$ and $a_{IJ}(\phi)$,  which can be combined into the
symmetric matrix
$${\cal N}_{IJ}= {1\over8}(a_{IJ}+i m_{IJ})\ .\eqn\nis$$
Note that
$(\Im{\cal N}_{IJ})$ is a positive definite $k\times k$
 matrix function. It will be useful to introduce the complex field
strengths
$$
{\cal F} ^{  I}_{\mu \nu} \equiv F^{  I}_{\mu \nu} - i\,
{}^\star F^{  I}_{\mu
\nu}, \qquad
{}^\star F^{\mu \nu  I} \equiv {1\over 2\sqrt{-h}}
 \varepsilon^{\mu\nu\rho\sigma}  F_{\rho\sigma }^I
\eqn\fstris$$
and rewrite the lagrangian as
$$
L=
-\,
\sqrt{-h} \, \Im \left({\cal N}_{IJ}{\cal F} ^{  I}_{\mu \nu}{\cal F} ^{
I {\mu \nu}}\right) \ .
\eqn\vecact$$
Introducing
$$G^{\mu \nu }_I = {\cal N}_{IJ}{\cal F} ^{ J \mu \nu}= -{i\over
\sqrt{-h}}{\delta S \over
\delta {\cal F} ^{  I}_{\mu
\nu}}\ ,
\eqn\gis$$
the $A_\mu^I$ field equations and Bianchi identities
can be written in terms of the $2k$-vector of two-forms
$$
{\bf F} = \pmatrix{ {\cal F}^I\cr G_I}
\eqn\athree
$$
as simply
$$\Im(d{\bf F})=0 \ .
\eqn\feld$$
These are formally preserved under the general linear transformations
$$\pmatrix{ {\cal F} \cr G } \to \pmatrix{ {\cal F}' \cr G '} =
\pmatrix{A& B \cr C &D}\pmatrix{ {\cal F} \cr G }\ ,
\eqn\vectran$$
where $A,B,C,D$ are arbitrary constant $k\times k$ matrices.
However, $G$ is not independent of $\F$, and these transformations must be
restricted to those
preserving the constraint relating $G$ to $\F$, and this will require
accompanying \vectran\ with a
transformation of $\N _{IJ}$:
$$\N _{IJ}(\phi) \to { \N '}_{IJ}(\phi)\ .\eqn\ntran$$
 Using
$G={\cal N} {\cal F}$, we obtain
$$\eqalign{G' &=C{\cal F}+DG=(C+D \N )\F
\cr
\F ' &=A\F +BG =(A+B\N) \F\ ,
\cr}
\eqn\fgpr$$
from which we obtain
$$G' = \N ' \F'\ ,
\eqn\gpris$$
where
$$\N ' (\phi) \equiv [C+D \N (\phi)].[A+B\N (\phi)] ^{-1}\ .
\eqn\npris$$
Moreover,   ${\N '}_{IJ}$ must be symmetric,
which together with \gpris,\npris, implies that the transformations
\vectran\ must be restricted to those preserving
the $2k\times 2k$ matrix
$$
\Omega = \pmatrix{0& \II \cr -\II &0}\ ,
\eqn\afour
$$
so that
$$M^t \Omega M =\Omega, \qquad M \equiv \pmatrix{A& B \cr C &D} \ .
\eqn\msym$$
Thus the group of such duality transformations \vectran,\ntran\ preserving the
field equations and
Bianchi identities is
$ Sp(2k;\R)$ [\dual,\GZ,\CDAFVP].

The matrix-valued function $\N _{IJ}(\phi) $ represents an (in general
infinite) set of coupling
constants, in the same way that the sigma-model metric did. This can be made
explicit if, for
example, $\N _{IJ}(\phi) $ has a Taylor series expansion about $\fii =0$,
$$\N _{IJ}(\phi) =
\N _{IJ,i}^{(0)}
+\N _{IJ,ij}^{(1)}\fii
+\N _{IJ}^{(2)}\fii \fj
+\N _{IJ,ijk}^{(3)}\fii \fj \fk
+\dots
\eqn\nexp$$
for some constants $\N _{IJ}^{(n)}$.
Substituting this in \vecact, we see that $\N _{IJ}^{(n)}$ is the coupling
constant for an
interaction of the form $\phi ^n \F \F$.
The $ Sp(2k;\R)$ symmetry of the equations of motion involves the
transformation \ntran\ of
$\N(\phi)$ and hence of the coupling constants, and so is a pseudo-symmetry in
general.

Note that the scalar fields $\phi$ are unchanged by the above action of $
Sp(2k;\R)$.
We can in addition consider scalar field transformations (i.e. sigma-model
target-space
diffeomorphisms)
$$ \fii \to \tilde \fii (\phi) \ .
\eqn\sdiff$$
The vector field action \aone\ will be left invariant provided this is
accompanied
by the transformation of the coupling constant generating function
$$\N _{IJ}(\phi) \to \tilde \N _{IJ}(\phi)\ ,
\eqn\ntransti$$
where
$$ \tilde \N _{IJ}(\tilde \phi(\phi)) =  \N _{IJ}(\phi) \ .
\eqn\ntiis$$
This is again a pseudo-symmetry in general.

Finally, one can combine a particular $Sp(2k;\R)$ transformation with a
particular diffeomorphism,
to obtain a pseudo-symmetry under which $\N$ transforms as
$$\N (\phi)\to \tilde \N '(\phi)\ ,
\eqn\tottran$$
where
$$ \tilde \N '(\tilde \phi(\phi)) =  \N '(\phi)
\eqn\ntipris$$
and $\N'$ is given in \npris.
However, it is sometimes possible to choose the diffeomorphism in such a way
that the change in
$\N$ it causes is such that it precisely compensates that due to the
$Sp(2k;\R)$
pseudo-duality transformation, so
that $\N$ is invariant:
$$\tilde \N '(\phi)= \N (\phi)     \ .
\eqn\niso$$
In such cases, the combined duality and diffeomorphism is a proper symmetry of
the theory.

\chapter{Pseudo-Duality Symmetries of Supergravity and Superstring Theories}

Consider now the system given by adding the scalar and vector lagrangians
\sig\
and \aone.
We have seen that it has a classical  pseudo-duality symmetry
$$D_{pseudo}=Diff(\cM) \times Sp(2k;\R)\ ,\eqn\pseud$$
and that the proper symmetries of the scalar sector alone consists of
the
isometry group $Iso(\cM)$.
In general, the group $D_{prop}$ of proper
symmetries is a subgroup of $Iso(\cM)\times
Sp(2k;\R)$.
In the theories of interest here (supergravities and  superstrings), the
duality group $Sp(2k;\R)$ contains a copy of the isometry group, and the group
$D_{prop}$ of proper
symmetries is a diagonal subgroup of these two copies
of $Iso(\cM)$:
$$D_{prop}=Iso(\cM)  \subset Iso(\cM)\times Iso(\cM) \subset Iso(\cM)\times
Sp(2k;\R) \subset D_{pseudo}\ ,
\eqn\embed$$
where $D_{pseudo}$ is given in \pseud.

In many cases, including many supergravity
theories, and all those with $N\ge  4$ supersymmetry, the scalar manifold
$\cM$
is a homogeneous
space
$G/H$ where
$H$ is the maximal compact subgroup of $G$.  The isometry
group
is then $G$ and for supergravity theories this is
indeed a subgroup of $Sp(2k;\R) $, so that \embed\ becomes \foot{Note that
$G/H$ sigma-models
can be formulated in such a way that in addition to a  rigid $G$ symmetry,
which is made manifest
by introducing extra scalars into the theory, there is   a local $H$
symmetry which can
be used to remove these extra scalars again. This local $H$ symmetry, which is
useful in coupling
to fermions, will not be considered here, but will be discussed in [\GH].}
$$G=D_{prop}  \subset G\times G \subset G\times
Sp(2k;\R) \subset D_{pseudo}=Diff(G/H) \times Sp(2k;\R)\ .
\eqn\embeda$$
For   $N=8$ supergravity,
  $k=28$, $G= E_{7(7)}$ and
$H=SU(8)$, so that
$\cM$ is the coset space $E_7/SU(8)$ with isometry group $E_7$,
 and the known proper symmetry group is $E_7$, while the full group of
pseudo-symmetries is
$$G_{pseudo}=Diff(E_7/SU(8)) \times Sp(56;\R)\ .\eqn\pseude$$
For $N=4$ supergravity
coupled to $m$ vector multiplets $k=6+m$, $G= SL(2;\R)\times O(6,m)$ and $H=
U(1)\times O(6)\times O(m)$, the known proper duality group is $SL(2;\R)\times
O(6,m)$, while the
pseudo-duality group is
$$G_{pseudo}=Diff[SL(2;\R)\times O(6,m)/U(1)\times O(6)\times O(m)] \times
Sp(56;\R)\ .\eqn\pseudf$$
For the general coupling of $N=2$ supermatter to $N=2 $ supergravity
with $m$ vector multiplets (so that $k=m+1$) [\gencouplN] it was found
that there is an $Sp(2k; \R) $   pseudo-symmetry [\CecFerGir]
  which contain the proper symmetries as a subgroup
[\dWVP] that   acts through isometries of the scalar
manifold $\cM$.
The $Sp(2k; \R) $   pseudo-symmetry has
 recently lead to interesting applications [\CDAFVP]. However,
this is a subgroup of an even larger pseudo-symmetry group, given by
\pseud.

The above applies, strictly speaking, only to the
bosonic truncation of the supergravity theories. However, using an analysis
similar to that of
[\GZ], (see also [\CDAFVP]) it can be seen that the symmetries found
above extend to pseudo-duality symmetries of the full supergravity
lagrangians.

For most such theories,     quantum effects break the vector pseudo-duality
group $Sp(2k;\R)$
down to the discrete subgroup $Sp(2k;\Z)$; this can be seen  using the
argument given in [\HT].
The $k$ abelian vector fields have $k$ electric charges and $k$ magnetic ones
which fit into the
{\bf 2k} representation of $Sp(2k;\R)$. If all $2k$ types of charge occur in
the theory, then the
Dirac-Schwinger-Zwanziger (DSZ) quantization condition implies that the
charges
lie on a
$2k$-dimensional lattice, and  $Sp(2k;\R)$ is broken down to the subgroup that
preserves the
lattice, which is $Sp(2k;\Z)$. It then follows from the embeddings \embed\
that
the proper duality
group is then also broken to a discrete subgroup, as was found in
[\Sen,\SS,\HT].
The $Diff(\cM)$ pseudo-duality symmetry can also be broken by quantum effects.
For example, if the sigma-model is coupled to chiral fermions, there can be
anomalies in the
sigma-model symmetry [\sigmodsymanom]; these can always be cancelled at the
expense of
adding a $D$-dimensional
Wess-Zumino term to the sigma-model and invoking the cancellation mechanism of
[\HW]. This and other
anomalies in the theory can lead to the breaking of some of the sigma-model
symmetries to
discrete subgroups, as will  be discussed in [\GNY,\GH].

We turn now to  the superstring. Arguments similar to those given in [\Sen,\SS]
for
S-duality and in
[\HT] for U-duality support the conjecture that discrete subgroups of the
pseudo-duality groups
\pseudf\ (with $m=22$ vector multiplets) and \pseude\ should be pseudo-duality
symmetries of
toroidally compactified heterotic string and type II string,
respectively. In each case, this
symmetry contains the group $Sp(56;\Z)$. These superstring pseudo-symmetries
and their implications
will be discussed further in [\GH].

\noindent{\bf Acknowledgements.}

A.V.P. wants to thank the Belgian Embassy in London for its support
through his appointment as `Visiting Belgian Professor in London'
and Queen Mary and Westfield College for  hospitality.
We wish to thank Bernard de Wit, Sergio Ferrara and Michael Green for
helpful discussions.

\refout
\bye